# Ballistocardiogram Artifact Reduction in Simultaneous EEG-fMRI using Deep Learning

James R. McIntosh, *Member, IEEE*, Jiaang Yao, Linbi Hong, Josef Faller,
and Paul Sajda, *Fellow, IEEE*

*Abstract*— *Objective:* The concurrent recording of electroencephalography (EEG) and functional magnetic resonance imaging (fMRI) is a technique that has received much attention due to its potential for combined high temporal and spatial resolution. However, the ballistocardiogram (BCG), a large-amplitude artifact caused by cardiac induced movement contaminates the EEG during EEG-fMRI recordings. Removal of BCG in software has generally made use of linear decompositions of the corrupted EEG. This is not ideal as the BCG signal is non-stationary and propagates in a manner which is non-linearly dependent on the electrocardiogram (ECG). In this paper, we present a novel method for BCG artifact suppression using recurrent neural networks (RNNs). *Methods:* EEG signals were recovered by training RNNs on the nonlinear mappings between ECG and the BCG corrupted EEG. We evaluated our model's performance against the commonly used Optimal Basis Set (OBS) method at the level of individual subjects, and investigated generalization across subjects. *Results:* We show that our algorithm can generate larger average power reduction of the BCG at critical frequencies, while simultaneously improving task relevant EEG based classification. *Conclusion:* The presented deep learning architecture can be used to reduce BCG related artifacts in EEG-fMRI recordings. *Significance:* We present a deep learning approach that can be used to suppress the BCG artifact in EEG-fMRI without the use of additional hardware. This method may have scope to be combined with current hardware methods, operate in real-time and be used for direct modeling of the BCG.

*Index Terms*— Electroencephalography (EEG), functional magnetic resonance imaging (fMRI), artifact removal, ballistocardiogram, deep learning, gated recurrent unit (GRU)

## I. INTRODUCTION

SIMULTANEOUS recording of electroencephalography (EEG) and functional magnetic resonance imaging (fMRI) presents a powerful approach for acquiring cortical data [1]. These simultaneous EEG-fMRI recordings hope to combine the high temporal resolution of EEG with the broad spatial sampling and superior spatial resolution of fMRI [2]. While challenges exist in fusing information from the two modalities (for example, see [3-6]), the technique has been widely used, for example in studies of basic human neuroscience studies of attentional orienting [7-9], perceptual decision making [10-12], value based decision making and reward processing [13-15] and analysis of resting state [16] as well as clinical applications such as epileptic event localization [17, 18].

Despite the advantages EEG-fMRI has to offer, one of the major difficulties of using the technique is the presence of various artifacts in the EEG data during concurrent fMRI-based image acquisition. One major artifact is known as the gradient artifact (GA), which arises due to the rapid switching of magnetic fields inside MRI scanners during fMRI recordings. Despite the large GA amplitude, it is possible to deal with these artifacts successfully through average artifact subtraction (AAS) [19], or optimal basis set (OBS) extensions since the GA repeats in a stereotypical way [20]. Another prominent artifact, which is our primary concern in this paper is known as the ballistocardiogram (BCG) [19, 21]. The BCG artifact arises due to the motion of EEG electrodes in the static magnetic field of the fMRI scanner. This motion is likely predominantly due to head movements during cardiac cycles [22] and local pulsatile movement of the scalp [23], both of which occur due to varying blood flow through scalp vessels during cardiac rhythms. The BCG artifact is characterized by complex spatiotemporal dynamics (Fig. 1), where its waveforms can vary greatly from channel to channel, as well as its intricate temporal relation with cardiac rhythms captured by the electrocardiogram (ECG).

Several previous studies have focused on BCG artifact removal, with one of the earliest proposed methods being AAS. However, unlike the GA, the BCG does not have a clearly time locked indicator. To deal with this problem, the ECG is pre-processed with a QRS complex detection method to produce a basis for the time locking, however, robust QRS complex detection is not trivial, and becomes increasingly difficult in high static field environments [23]. Additionally, the QRS

Manuscript submitted to arxiv.org Oct 14, 2019. This work was supported in part by the National Institute of Mental Health under grant R33MH106775, the United States Army Research Laboratory under Cooperative Agreement W911NF-10-2-0022 and a seed grant for MR Studies Program of the Zuckerman Mind Brain Behavior Institute at Columbia University (CU-ZI-MR-S-0006). The views and conclusions contained in this document are those of the authors and should not be interpreted as representing the official policies, either expressed or implied, of the United States government. *(James R. McIntosh and Jiaang Yao contributed equally to this work.) (Corresponding authors: James R. McIntosh; Jiaang Yao.)*

J. R. McIntosh and J. Yao are with the Department of Biomedical Engineering, Columbia University, New York, NY 10027, USA. (correspondence e-mails: j.mcintoshr@gmail.com; jy2951@columbia.edu)

L. Hong is with Department of Biomedical Engineering, Columbia University.

J. Faller is with the Department of Biomedical Engineering, Columbia University, and the Human Research and Engineering Directorate, U.S. Army Research Laboratory, Aberdeen Proving Ground, MD 21005, USA

P. Sajda is with the Department of Biomedical Engineering, and the Data Science Institute, Columbia University.

complex event time is itself not indicative of the corresponding peak in the BCG, so that to subtract around this point the BCG is often shifted by a fixed delay of 210 ms [19] (but see [24]). Ultimately, the weakness of AAS is that the BCG does not form clean repeats around the shifted QRS complex peak. To address this, another method developed both for GA and BCG is OBS [20]. In this algorithm, an AAS step is used in combination with principal component analysis (PCA) to retain signal components strongly related to the BCG, these are then subtracted in adaptive fashion from the corrupted signal. This algorithm may be an improvement over AAS, however, the method still relies on QRS complex detection and is dependent on the number of components marked as artifacts, which can potentially interfere with the interpretation of the underlying neural data. This is in part because BCG artifacts may not always be linearly separable from neural activity [24]. Other software based methods, for example based on independent component analysis (ICA) [25-27], or harmonic regression [28] have also been suggested, but it is not yet clear whether they provide systematic advantages over OBS.

Deep learning [29] has gained popularity in recent years, and its use in neuroscience and related fields has become pervasive. For example, a small selection of applications include generic task relevant classification based on EEG activity [30, 31], markerless motion tracking [32], modeling and characterizing neurons in visual cortex [33], modeling visual processing [34], modeling and analysis in psychiatric and neurological disorders [35], and neuron tracing and segmentation [36, 37].

In this paper, we propose a novel pipeline for BCG suppression using recurrent neural networks (RNN) [38, 39], specifically, gated recurrent units (GRU) [40] configured in a deep architecture. In assuming that cardiac rhythms do not contain significant cortical information, we train our network (termed BCGNet) on ECG and BCG corrupted EEG data, in order to uncover the nonlinear state dependent mapping between ECG and BCG signals. This mapping can then be subsequently used for subtracting the BCG waveforms from the EEG data. Our method generates robust and reliable prediction of BCG waveforms and cleaned EEG data, and importantly does not remove the signal reflecting the underlying neural activity, as it improves single-trial analysis of the EEG.

## II. METHODS

### A. Subjects and Behavioral Paradigm

A total of 25 subjects were scanned and six were excluded from further analysis. Two were rejected due to missing neuroimaging data, two were rejected due to abnormalities in the collected neuroimaging data, one was rejected due to excessive movement, and one was rejected due to drowsiness and inability to complete the task as per instruction. For the final 19 subjects included in this study, 6 were male and 13 were female. Age ranged from 18 to 32, with a mean of 25.9 and a standard deviation of 3.6 years. All participants had normal or corrected-to-normal vision and no history of psychiatric illness or head injury. The Columbia University Institutional Review Board (IRB) approved all experiments and informed consent was obtained before the start of each experiment.

An auditory oddball paradigm (see [41]), was performed by subjects while fMRI and EEG were simultaneously recorded. The paradigm included 80 % standard and 20 % of oddball (target) stimuli, where standard stimuli were pure tones with a frequency of 350 Hz, while the oddball stimuli were broadband (laser gun) sounds. Stimuli lasted for 200 ms with an inter-trial interval (ITI) sampled from a uniform distribution between 2 s and 3 s. Stimuli were presented through MR compatible earphones, and subjects were instructed to respond to oddball sounds as quickly and as accurately as possible, by pressing a button on an MR-compatible button box (PYKA, Current Designs, PA, USA). Subjects were told to ignore standard tones.

Every subject was scheduled to complete five runs in total (105 trials per run), with an average of 4.7 runs per subject (range of three to five, standard deviation of 0.7). At the processing stage, we further rejected a single run from three subjects as we could not perform either successful GA removal or QRS detection on them so that the final average number of runs per subject was 4.6 (range of two to five, standard deviation of 0.98).

### B. Simultaneous EEG and fMRI Data Acquisition

EEG and fMRI were recorded inside a 3T Siemens Prisma scanner using a 64 channel head coil. During the task, functional echo planar imaging (EPI) data were collected with 3 mm in-plane resolution and 3 mm slice thickness. 42 slices of 64 x 64 voxels were acquired using a 2100 ms repetition time (TR) and 25 ms echo time (TE). EEG was recorded with a 64 channel BrainAmp MR Plus system (Brain Products, Germany), at a sampling rate of 5 kHz. The 64 channels include

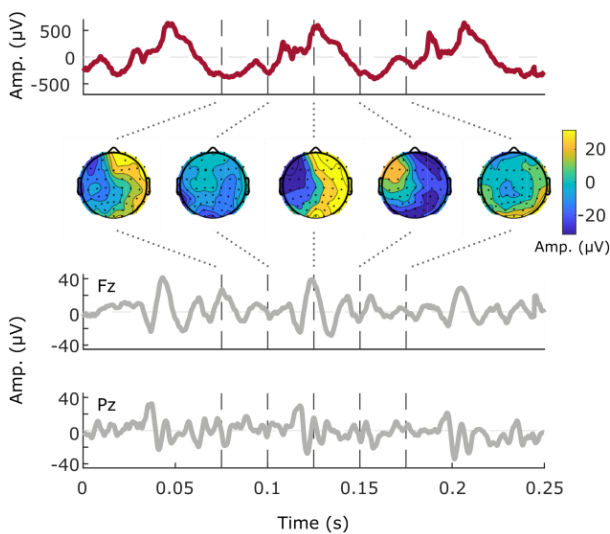

Fig. 1. Example ECG and accompanying BCG corrupted EEG. Top: Example ECG trace for three cardiac cycles. Middle: Corresponding corrupted EEG topography at marked times in the cardiac cycle. Bottom: Example BCG trace over the corresponding time for channel Fz (top) and Pz (bottom), showing the diverse temporally shifted non-linear transformation from ECG to BCG.

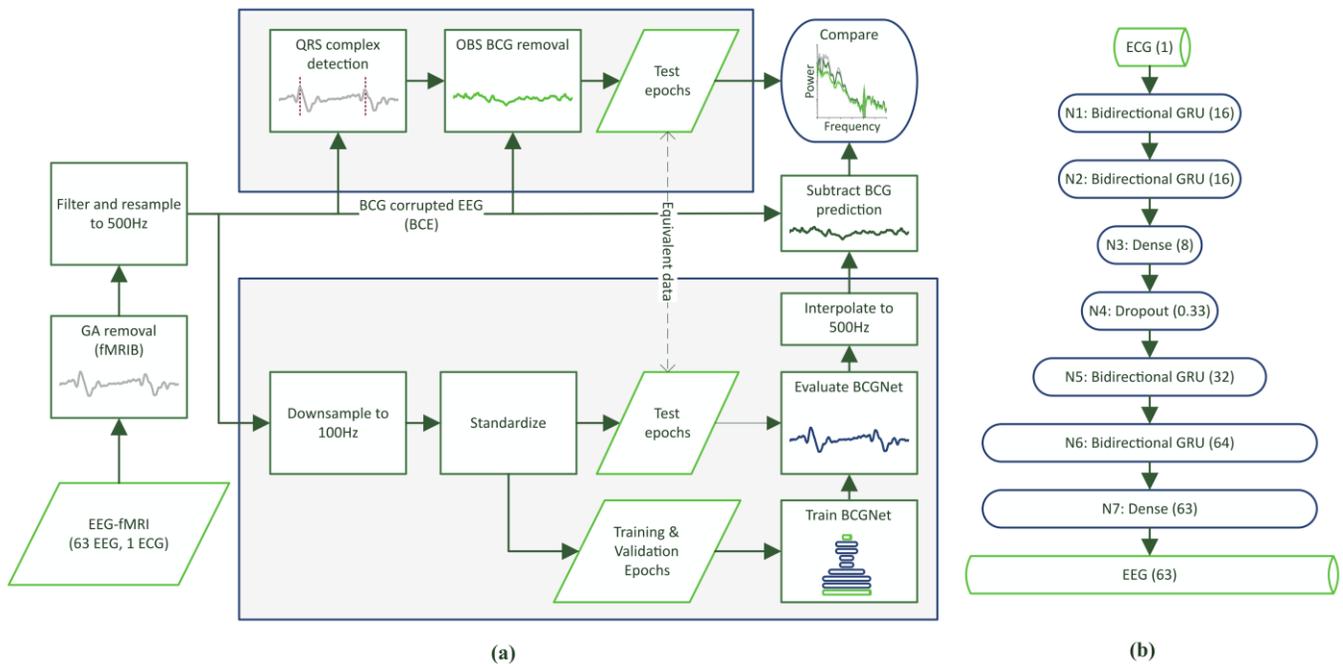

Fig. 2. Processing pipeline and network architecture. (**a**) Major processing steps performed on the raw data to generate OBS and BCGNet cleaned datasets. After GA removal and resampling, data is processed independently by the OBS method and by BCGNet. Equivalent test epochs that were not used for BCGNet training are then compared between the two methods. Note that the comparison step refers only to our power spectral density measures (Fig. 4), as the evoked response and classification based analyses use all the data, not just the matched test set epochs. (**b**) Standard BCGNet architecture with layer sizes in parentheses. Additional L2 regularization parameters were also used (recurrent: N1 = 0.096, N2 = 0.090, N5 = 0.024, N6 = $2.5 \times 10^{-7}$; activation: N1 = 0.030, N2 = 0.013, N5 = 0.067, N6 = 0.055).

63 cap electrodes and 1 ECG electrode in an extended 10-20 configuration with ground electrode at AFz and reference electrode at FCz. Inside the scanner, we used sandbags to stabilize the amplifiers against any potential vibrations caused by the scanner. To avoid amplifier saturation and ensure high SNR, we used the following settings in the Brain Products Recorder environment. 1) Voltage resolution at 0.5 μV, 2) high-pass cut-off frequency at 0.1 Hz, 3) low-pass cut-off frequency at 250 Hz and 4) all electrode impedances controlled to be under 20 kΩ.

### C. EEG Pre-processing

Raw data was initially imported into MATLAB using the EEGLAB toolbox [42]. We subsequently low-pass filtered with a non-causal finite impulse response (FIR) filter, cutoff at 70 Hz and then performed GA removal on the dataset using FASTR [43], part of the FMRIB plugin for EEGLAB, provided by the University of Oxford Centre for Functional MRI of the Brain (FMRIB) [20, 44].

The data was resampled to 500 Hz, and non-causally high-pass filtered at 0.25 Hz to reduce electrode drift, while not interfering with cardiac related information. We will refer to the output of this processing step as "BCG corrupted EEG" (BCE, see Fig. 2a).

### D. BCGNet based BCG Suppression

In our proposed method, we made use of the MNE package [45, 46] to load our BCG corrupted EEG into Python (v3.7). Data was further resampled to 100 Hz due to computational constraints, and each channel was normalized to zero-mean and unit variance within each run. After BCG signal prediction was carried out, the downsampling step was reversed by using a Piecewise Cubic Hermite Interpolating Polynomial (PCHIP) [47] to upsample our predicted BCG signal so that it could be subtracted from our BCG corrupted EEG. Data was then epoched into 3 s consecutive chunks. We were concerned that epochs with very large artifacts may be detrimental for our training procedure, and we consequently took the magnitude of the epochs, averaged over channels and time and used these values to calculate the median absolute deviation. Any epochs corresponding to a value greater than five times the median absolute deviation was then rejected (this resulted in 2.7 % of epochs being rejected). Epochs are then randomly assigned to a training set (70 %), validation set (15 %) or test set (15 %).

### E. Training

While several architectures were tested, our general approach was to stack regularized GRU layers, mapping the ECG to the BCG corrupted EEG, through a final linear dense layer. Specifically, our network takes in 1-channel ECG data and outputs 63-channel EEG data. Our loss function was designed to minimize the mean squared error (MSE) with the Adam optimizer [48]. Training was done on an NVIDIA GeForce RTX 2080 and Tesla P100 GPUs with CUDA 10.0 and cuDNN v7.1, in Tensorflow v1.14 [49], using the Keras API [50], and CuDNNGRU layers were used to speed up training. Early stopping based on the validation set using a liberal patience of 25 epochs was used, allowing selection of model weights that generate the lowest MSE on the validation set.

## F. Architecture Selection

In order to explore the large space of neural network models that may be suitable for BCG prediction from ECG, we began by applying a simple architecture on run one of all subjects. The architecture we used consisted of three hidden bidirectional layers, each with 16 GRU cells, followed by a dense layer with a linear activation function. We then sub-selected three subjects that span the range of test set MSE (one subject from each of the lowest, middle and highest quintile). We subsequently used Hyperas [51], a Python package that combines Hyperopt [52] and Keras, for hyperparameter tuning on these three subjects. In particular, we were interested in knowing the number of hidden and non-hidden layers, the number of cells in each layer, as well as values for regularizing parameters. This approach enabled hyperparameter tuning and exploration of the space of architectures in a reasonable time. With the results of the Hyperas runs, we examined the patterns of the best performing architectures and used common properties to construct a more general architecture which is shown in Fig. 2b, which we refer to as BCGNet throughout this manuscript.

## G. Optimal Basis Set BCG Reduction

In order to generate a comparison for our proposed method, we used the FMRIB EEGLAB plugin to perform BCG artifact removal. The BCG corrupted EEG was high-pass filtered at 2.5 Hz, and QRS complex detection [53, 54] was performed. A separate copy of the BCG corrupted EEG (with the 2.5 Hz high-pass filtering step omitted) was then fed to the FMRIB plugin's BCG suppression function combined with the generated QRS complex event times. The BCG suppression function was run in OBS mode, with the number of bases set to four (the default).

## H. Validation

### 1) Power Spectral Density Measures

We estimated the power spectral density (PSD, Welch's method with a Hanning window) for all EEG channels in each test epoch for BCGNet, OBS and the BCG corrupted EEG. In order to compare the performance of the two methods under investigation, we then calculated the ratio of the PSD for BCGNet and OBS against the BCG corrupted EEG signal from its PSD. This provides us with a measure of power reduction of the two BCG suppression techniques for all frequencies, runs and channels. The median was then calculated across the frequency band of interest (either delta 0.5 to 4 Hz, theta 4 to 8 Hz, or alpha 8 to 13 Hz), test epochs and channels of interest (including the average of all channels), and this quantity was then averaged with an arithmetic mean across runs. A Wilcoxon signed-rank test across subjects, between methods was then carried out on the channels and bands of interest and Bonferroni-Holm corrected across frequency bands for Fz, Pz and the channel average independently.

### 2) Evoked Response Analysis

While the PSD is generally informative about whether the BCG signal is being removed, it does not discriminate between suppression of BCG (or other source of artifact), and suppression of signal reflecting the underlying neural activity. In an attempt to constrain our analysis to the outcome of BCG suppression on this signal, we evaluated our within subject models on the entire ECG time series for each run. We then compared the OBS method with our BCGNet method after current source density estimation (CSD) [55], applying a further 0.5 Hz high-pass and 50 Hz low-pass [56]. Data was epoched into sections from -0.5 s to +1.5 s relative to standard and target stimuli.

In BCG corrupted EEG, the across trial standard deviation of evoked responses is likely to be large and dominated by different time portions of the BCG. It follows that one way to assess the quality of the BCG suppression is to investigate how the standard deviation across trials is reduced for the methods under test. To do this, we calculate the time-locked and time-resolved standard deviation across standard trials, and separately target trials for each subject and each channel of interest. In order to compare this measure across methods we averaged it from 0.25 s to 0.75 s post stimulus and performed a Wilcoxon signed-rank test across subjects. Bonferroni-Holm correction was applied across comparisons for the displayed channels, Fz and Pz, as well as the channel average independently.

### 3) Task Relevant Single-Trial Classification

Another method to compare BCG suppression techniques which gets at the heart of whether we are disrupting task-relevant EEG signal is to build a classifier that discriminates between targets and standard stimuli. In order to do this, we initially filtered and epoched the data as in our evoked response analysis. We then used logistic regression with L2 regularization [57] applied over features composed from all samples from all channels within 60 ms sliding windows moving in 20 ms steps (see [41, 56] for more details on this approach) to estimate the area under the receiver operator characteristic curve (AUC) as a measure of performance. AUC was averaged from the results of six fold cross-validation, with a further internal six fold cross validation to determine the regularization parameter. The regularization parameter was selected by following the "one-standard error" rule, where the largest value for the parameter is used such that the deviance is within one standard error of the minimum [58]. In order to compare classification performance for different BCG suppression methods AUC was averaged from 0.25 s to 0.75 s post stimulus, and a Wilcoxon signed-rank test across subjects was then carried out and Bonferroni-Holm corrected.

### 4) Across Subject BCGNet Training

After evaluating our model, and comparing its performance to the OBS method, we proceeded to train it on all training data from all subjects while holding one subject out entirely (this was done for each subject). Evaluation was then done on the held-out subject, followed by an evaluation on the test data set of the held-out subject after further training on their data set. This procedure enabled us to examine how well our model generalizes to new subjects, and how quickly our model can be trained on a new subject's data if it has been pre-trained.

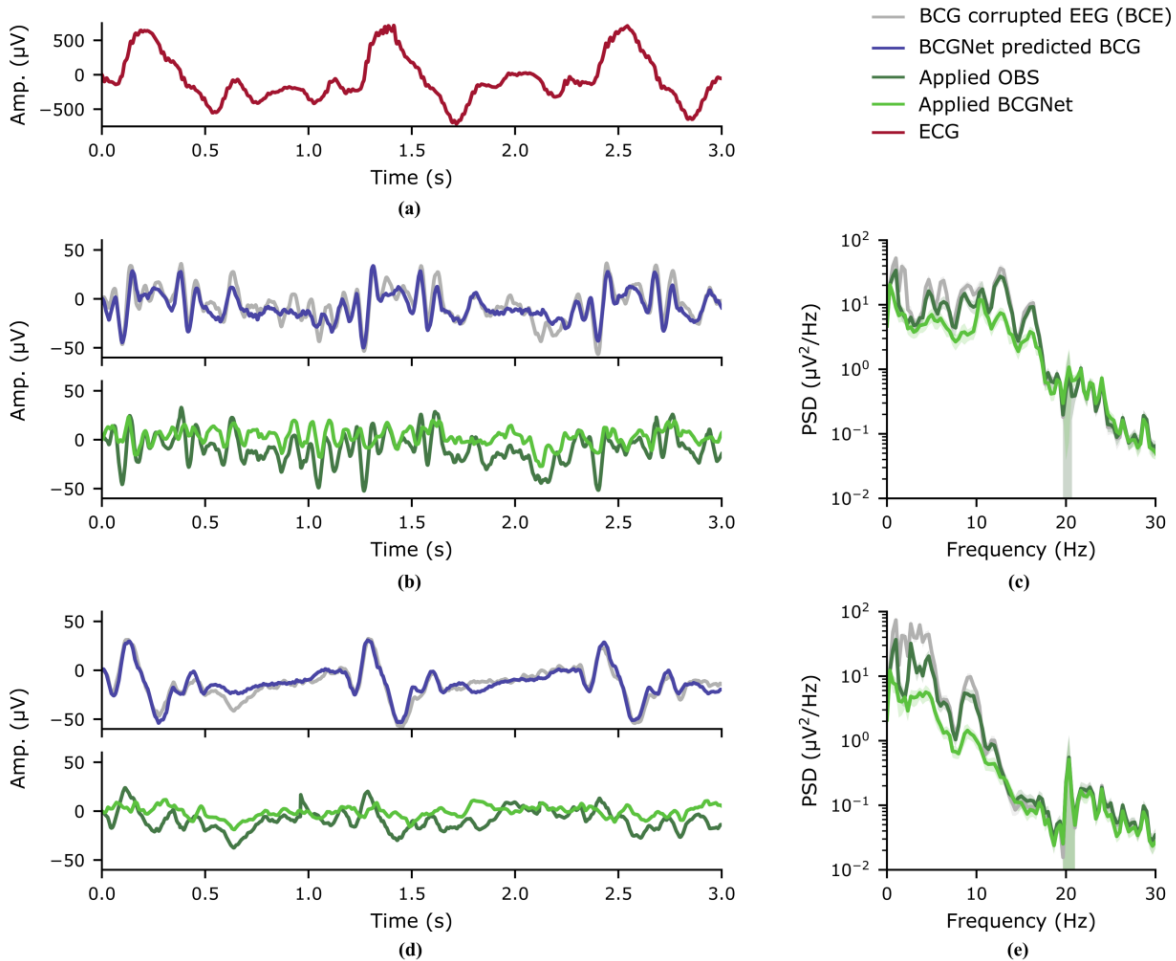

Fig. 3. Comparison of BCGNet and OBS methods for BCG suppression in a single subject. (**a**) ECG over a selected time period. (**b**) Top: BCGNet predicted BCG for channel Fz in a test epoch (blue), overlaid on gradient artifact removed BCG corrupted EEG (BCE) signal (gray). Bottom: Corrupted EEG with subtracted BCGNet predicted BCG (light green) as well as OBS cleaned EEG for comparison (dark green). (**c**) Power spectral density of corrupted EEG, BCGNet cleaned EEG, and OBS cleaned EEG for channel Fz. OBS shows a reduction in power over the corrupted EEG signal as expected, but BCGNet shows substantial further reduction. Shaded regions show 2 × SEM. Note that the large variance at 20 Hz is due to residual GA. (**d**) Same as (b) but for channel Pz. (**e**) Same as (c) but for channel Pz.

## III. RESULTS

### A. BCGNet Models BCG Signal when Trained Within-subject

We initially compare the performance of OBS and BCGNet at suppressing BCG signal power by examining the PSD in test epochs. We show a single subject example for channels Fz and Pz in Fig. 3, demonstrating that BCGNet can predict the progression of the BCG time series despite its inhomogeneous behavior across channels.

Fig. 4 shows the power ratio divided by EEG bands for all subjects calculated from the test epochs. Interestingly when considering an average over channels, it appears that for low frequencies (delta) OBS is approximately as successful as BCGNet at power reduction with a median power ratio difference (MD) of -0.03 ($p = 0.334$), with larger differences in specific channels (e.g. in Fz, MD = -0.05, $p = 0.044$). On the other hand, BCGNet is clearly more successful in theta (MD = 0.06, $p = 0.003$) and alpha bands (MD = 0.10, $p = 4 \times 10^{-4}$). We were concerned that part of the reason why BCGNet does not outperform OBS in the delta band (and in fact displays higher power) is because OBS is too aggressive and removes EEG signal. This is a lesser concern for BCGNet as the network is not able to directly adapt to the BCG corrupted EEG test epochs, and we investigate this in the following sections.

### B. Relation to QRS Complex Detection

One reason for BCGNet performance generally exceeding that of OBS, may be due to the QRS complex detection, since OBS relies on detection of all QRS complexes. In order to investigate this, we calculated the power ratio of the alpha band power (as for Fig. 4), where the relative power reduction of BCGNet is most striking. We then modeled the change of this ratio between OBS and BCGNet as linearly dependent on the $\Delta$QRS standard deviation. The $\Delta$QRS standard deviation was used as a measure of QRS detection quality. We found (see Fig. 5) a relation between the fraction of outlier $\Delta$QRS and the change in power ratio (F-test vs constant model, $p = 0.015$, n = 19), suggesting that indeed BCGNet is relatively robust to cases where OBS fails due to poor QRS detection. It should be noted however, that this effect appears to be driven by a small subset of subjects, and that BCGNet generally reduces alpha band power more than OBS for most subjects.

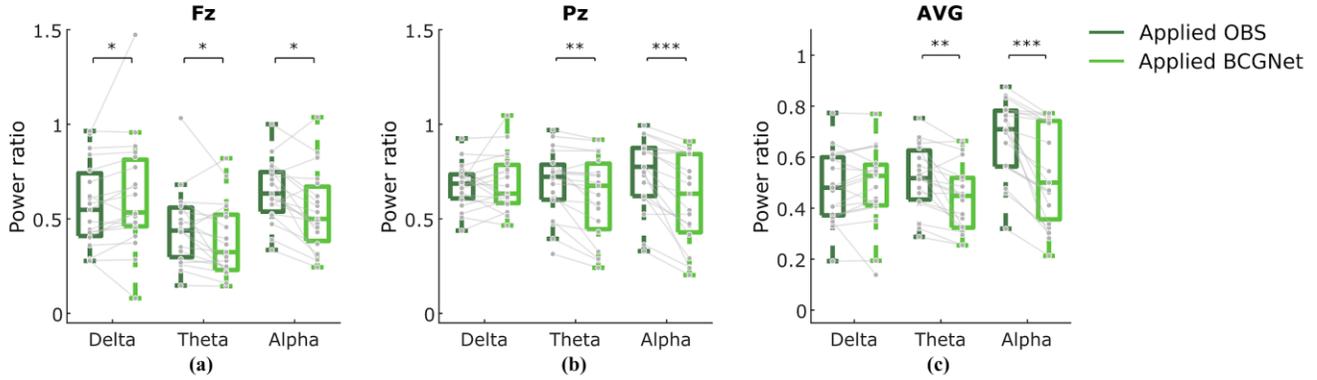

Fig. 4. Comparison of BCGNet and OBS methods for BCG suppression for all subjects. (**a**) Boxplot of average power ratio in relevant EEG bands for OBS and BCGNet methods compared to BCG corrupted EEG (BCE) signal for channel Fz. Linked grey dots represent individual subjects, median differences (MD) are presented as the median of the OBS – BCGNet difference. Hinges represent first and third quartile and whiskers span the range of the data not considered outliers (approximately $\pm 2.7\sigma$). Both methods strongly reduce signal power, with OBS appearing more effective in the delta band while BCGNet appears more effective in the theta and alpha band. (delta MD = -0.05, $p = 0.044$; theta MD = 0.04, $p = 0.044$; alpha MD = 0.11, $p = 0.024$). (**b**) Same as (a), but for the channel Pz (delta MD = -0.01, $p = 0.717$; theta MD = 0.03, $p = 0.002$; alpha MD = 0.11, $p = 6 \times 10^{-4}$). BCGNet reduces power more effectively in the theta and alpha bands. (**c**) Same as (a), but for the across channel average. BCGNet power reduction is significantly larger than for OBS in the theta and alpha band (delta MD = -0.03, $p = 0.334$; theta MD = 0.06, $p = 0.003$; alpha MD = 0.10, $p = 4 \times 10^{-4}$). OBS more effectively reduces power in the delta band than BCGNet, although this difference does not appear to be statistically significant.

### C. BCGNet Improves Task Relevant Metrics

Because power reductions in the BCG corrupted EEG signal can have multiple sources, we further compare OBS and BCGNet using task relevant metrics. We expect that if BCGNet is able to reduce BCG more effectively than OBS (rather than simply signal power), that we should see a reduction in standard deviation of the evoked responses. Indeed, as shown in Fig. 6 this is what we find. In particular, when examining the average standard deviation across channels for targets, we find that OBS reduces evoked standard deviation over the gradient artifact removed BCG corrupted EEG (BCE) signal ($p = 8 \times 10^{-4}$, Wilcoxon signed-rank test), and in turn BCGNet reduces the evoked standard deviation over OBS ($p = 9 \times 10^{-4}$) and the BCE signal ($p = 5 \times 10^{-4}$). This is broadly replicated for specific channels and for both standards and targets as shown in Fig. 6.

A yet more direct method to examine how OBS and BCGNet impact the EEG signal in a task relevant manner is to attempt to use the cleaned signal for classification. As shown in Fig. 7, and consistent with our hypothesis, we find that BCGNet increases classification accuracy over the BCE signal ($p = 4 \times 10^{-4}$). We found a similar pattern of results for band amplitude based classification (see supplementary materials). Interestingly, we also see a low p-value ($p = 0.053$) that OBS decreases classification accuracy indicating that it reduces task relevant SNR.

### D. Across Subject Generalization

We performed a 'leave-one-subject-out' experiment and show that the network struggles to generalize in its current form. This is no surprise, as there is substantial variability across subjects in the properties of the BCG. Upon retraining our model on the training set of the held-out subject, our power suppression and classification performance exhibit the same general pattern when compared to OBS (Fig. 8a-b). When compared to direct within subject training however, the re-trained network power-ratio is significantly higher (Supp. Fig. 2a). The main advantage of using a pre-trained network is that it reduces further computational cost, in this case cutting the median number of training epochs from 59 to 10 (Supp. Fig. 2b).

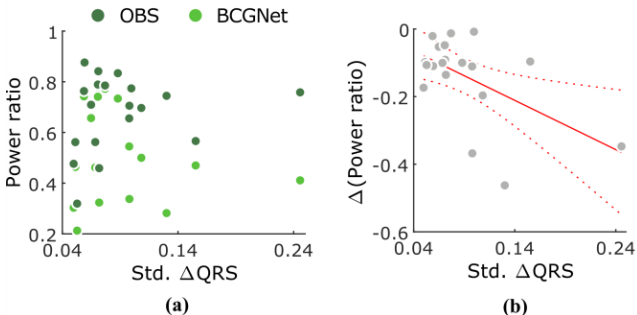

Fig. 5. BCGNet reduces BCG signal power in the alpha band more effectively than OBS regardless of QRS complex detection quality, however the improvement may be more pronounced when QRS detection quality is worse. (**a**) Ratio of cleaned EEG to corrupted EEG in the alpha band (8-13Hz) averaged over all channels for the BCGNet (light green) and OBS (dark green) method. (**b**) Difference in power ratio between BCGNet and OBS from (a) plotted against $\Delta$QRS standard deviation. Solid red line shows fit (F-test vs constant model, $p = 0.015$, n = 19) indicating a relation between BCGNet improvement over OBS and QRS detection quality. Dashed lines correspond to 95 % confidence intervals.

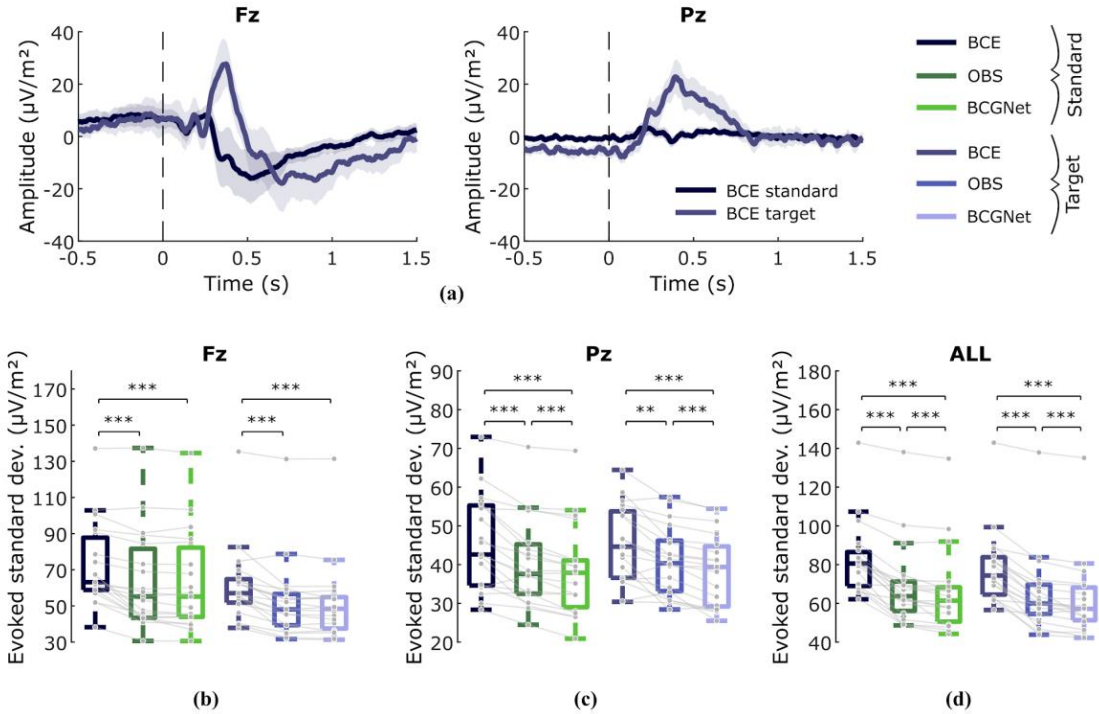

Fig. 6. BCGNet does not alter the evoked response when compared to OBS, however standard deviation of evoked response is reduced. (**a**) Mean across subject evoked response to stimulus (standard or target presented at 0 s) for gradient artifact removed BCG corrupted EEG (BCE) signal for task relevant Fz and Pz channels. Shaded region represents 2 x SE across subjects. BCGNet and OBS methods not shown due to the similarity to the BCE mean evoked potential. (**b**) Boxplot of standard deviation of evoked responses in the task relevant regions from 0.25 s to 0.75 s post stimulus for individual subjects. Linked grey dots represent individual subjects, for Fz we found no significant difference between BCGNet and OBS for standards (st) or targets (ta), although the difference was present between either method and the BCE signal (BCE/OBS st: $p = 9 \times 10^{-4}$, ta: $p = 9 \times 10^{-4}$; BCE/BCGNet st: $p = 9 \times 10^{-4}$, ta: $p = 9 \times 10^{-4}$; OBS/BCGNet st: $p = 0.117$, ta: $p = 0.421$). Hinges represent first and third quartile and whiskers span the range of the data not considered outliers (approximately ±2.7 σ). (**c**) Boxplots as in (b) but for channel Pz. Here, both methods reduce task relevant standard deviation relative to BCE, although BCGNet further reduces task relevant standard deviation of the evoked responses relative to OBS (BCE/OBS st: $p = 9 \times 10^{-4}$, ta: p = 0.002; BCE/BCGNet st: $p = 8 \times 10^{-4}$, ta: $p = 8 \times 10^{-4}$; OBS/BCGNet st: $p = 9 \times 10^{-4}$, ta: $p = 7 \times 10^{-4}$). (**d**) Boxplots as in (b) for the standard deviation averaged across channels. As in (c), standard deviation is reduced by OBS from the BCE signal, and further reduced by BCGNet (BCE/OBS st: $p = 8 \times 10^{-4}$, ta: $p = 8 \times 10^{-4}$; BCE/BCGNet st: $p = 7 \times 10^{-4}$, ta: $p = 5 \times 10^{-4}$; OBS/BCGNet st: $p = 9 \times 10^{-4}$, ta: $p = 9 \times 10^{-4}$).

## IV. DISCUSSION

We have shown that our method reduces power in standard EEG frequency bands, while simultaneously improving task relevant metrics over the conventional OBS method. Combined, these results are a clear indication that BCGNet suppresses BCG artifact more effectively than OBS. Furthermore, we have shown that a pre-trained network can substantially aid training when applied in a new subject, and acts to enable fast learning of the BCG structure.

A subtlety appears to be that in the delta band, suppression of power for BCGNet is not significantly higher than for OBS, and a general trend in the opposite direction could in fact be argued for. However, results of classification which are likely dominated by low frequency content, based on OBS and BCGNet cleaned signals, indicate better signal quality after processing with BCGNet. It could be argued that this occurs because BCGNet is relatively more effective at higher frequencies (theta and alpha bands) than OBS, however because classification accuracy based on the OBS cleaned signal shows a trend for being lower than accuracy on the gradient artifact removed BCG corrupted EEG signal, this may be in fact due to OBS suppressing signal reflecting underlying neural activity at low frequencies. Taken together, these results point to BCGNet being more successful at BCG signal reduction independent of band frequency.

### A. Limitations

In general, most BCG signal suppressing methods have more constrained models than what we propose in this work. However, the potential removal of EEG reflecting the underlying neural activity is not unique to our method, but also present when any attempt is made to construct BCG artifact with a subset of components extracted by relying on dimensionality reduction of the BCG corrupted EEG, such as ICA or PCA. However in principle, our method is particularly susceptible to specifically removing ECG related EEG signal in cases where it is present. One of our foundational assumptions is that the EEG is independent of the ECG. This is clearly not valid under all circumstances. For example, [59] discuss how the QRS complex timing is related to response times during a decision task, so it seems clear that the exact timing of the cardiac cycle is likely to have some influence on specific brain function. In fact AAS, and by extensions OBS and many proposed methods also make the assumption that ECG and EEG are uncorrelated [21]. However, because deep neural networks are particularly powerful, our method may remove EEG signal that is dependent on ECG that standards methods may be unable to accurately model. This is likely to be a relevant criticism, however with current experimental paradigms it seems safe to assume that in general, any ECG related activity in EEG should be treated as a confound (as its source is ultimately unverifiable), so that its removal is not detrimental. Plainly, EEG-fMRI may not currently be a good experimental setting for studying heart-brain relations. In our task, because classification improves and inter-trial evoked response standard deviation decreases it seems that BCGNet did not have a particular disadvantage because of its likely ability to remove cardiac related EEG activity. In contrast to this, OBS based classification performance was reduced when compared to BCG corrupted EEG based performance which suggests that the method may have detrimentally removed task relevant signal.

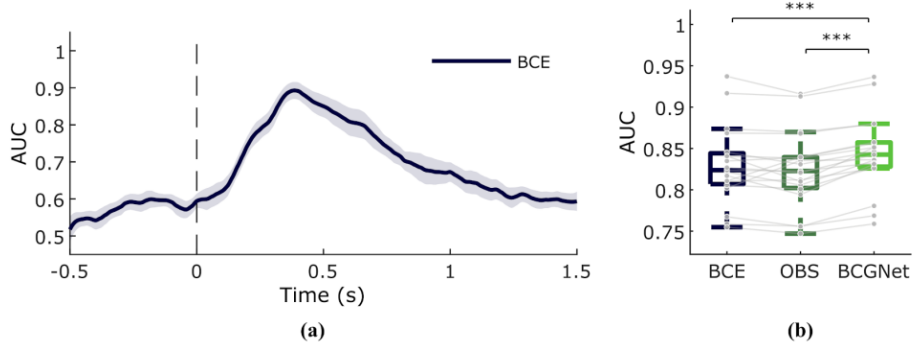

Fig. 7. Overall classification quality is modestly but consistently improved by application of BCGNet when compared to OBS. (**a**) Average across subject receiver operator characteristic area under the curve (AUC) corresponding to standard vs target sliding window classification (shaded region corresponds to 2 x SE across subjects) for BCG corrupted EEG (BCE) signal. (**b**) Boxplots of AUC averaged in the task relevant region from 0.25 s to 0.75 s post stimulus for the BCE and tested BCG suppression method. Hinges represent first and third quartile and whiskers span the range of the data ($MD_{BCE-OBS} = 0.004$, $p = 0.053$; $MD_{BCE-BCGNet} = -0.011$, $p = 4 \times 10^{-4}$; $MD_{OBS-BCGNet} = -0.015$, $p = 4 \times 10^{-4}$).

We hypothesize that the main limitation of our method implemented in its current form is that we could not feed our input stage with sufficient information regarding sources of artifact. In principle, it is possible that our method can address some movement of the head when it is coupled with a change of the ECG, as the change in cardiac rhythm state progression may be detected and used as an indicator of a change in BCG state progression. Essentially, while our model weights are currently fixed at evaluation time, the model itself is in principle adaptable to changing circumstances (such as changes in heart rate). However, active head motion that is unrelated to the ECG is undetectable to our method and will likely result in estimation errors of the BCG. Alternative approaches for BCG suppression that depend on hardware (rather than already discussed software methods) extensions that go beyond EEG and ECG appear increasingly promising. Most of these methods work by adding reference sensors to the head (or specific sources of noise), and then removing their contribution to the EEG signals in software with varying degrees of sophistication [60-65]. While the reference layer often may not provide a perfect mirror for the BCG signal, and in general hardware solutions involve a certain overhead, methods that exploit reference signals are likely to provide an edge in the artifact removal process. We conjecture that our proposed method could be used to enhance the effectiveness of current reference layer methods, and address what we perceive to be the main limitation of our method - a lack of information regarding head movement and other sources of noise.

Another limitation of our current study is that we have not compared our results to data recorded outside of the range of the static magnetic field. Indeed, a specifically designed paradigm involving in and out of scanner experiments under different noise conditions would be ideal, and data from such recordings may enable generative adversarial extensions of our method that could prove to be yet more powerful.

*B. Advantages*

We believe that in finding the direct ECG to BCG mapping, we have identified a novel approach to BCG suppression, which may be further refined with a larger body of data.

The fact that our approach does not rely on summarizing the cardiac cycle in terms of discretely detectable QRS complex events appears to be one of the advantages of our method. It may be argued that the comparison between BCGNet and OBS is not entirely fair because OBS relies on QRS complex detection, so that if the QRS complex detection is poor OBS is at a disadvantage. While this is in part the point, as we do not get to choose whether or not any individual subjects has good or bad quality QRS complex detection, we have also shown that avoiding QRS complex detection is not the only advantage conferred by BCGNet. It has also been observed [23], that QRS complex detection generally becomes more difficult at high field strengths, making a lack of reliance on the method increasingly relevant.

A more subtle advantage of our method is that it remains applicable in unusual experimental settings. For example, in EEG-fMRI-TMS [66, 67] a TMS pulse might cause a small QRS complex like artifact in the ECG. Since the timing of these occurrences is known, a straight-forward extension of our method can allow us to continue running our model based on its own predictions of the upcoming ECG. At the same time, the EEG itself is corrupted much more strongly, making standard BCG suppression difficult, however our forward model can continue being evaluated enabling recovery of the EEG signal that only depends on how quickly it recovers from the TMS pulse.

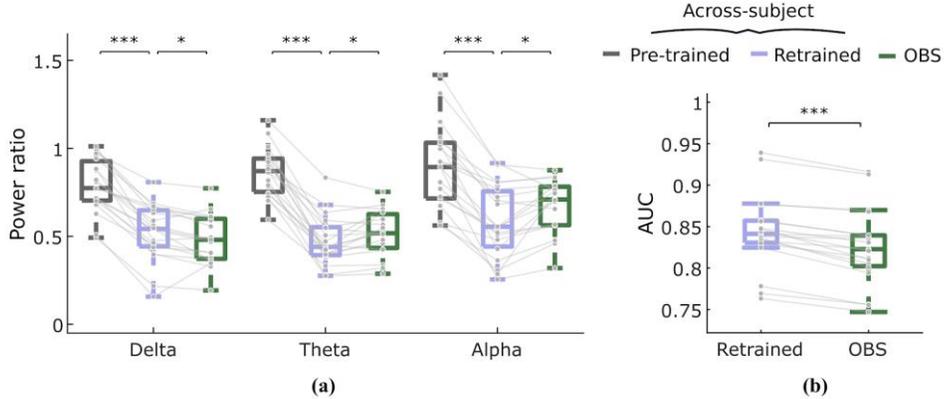

Fig. 8. Transfer in network trained across subjects. (**a**) Power ratio for BCGNet trained on all subjects and evaluated on held-out subjects for average channel (gray, pre-trained, PT) demonstrates poor performance which is improved upon substantially by within subject retraining (light blue, retrained, RT). OBS method (dark green) as in Fig. 4c plotted for comparison. Bonferroni-Holm correction was carried out across bands and networks, although we limited our tests to PT compared to RE (delta $MD_{PT-RT} = 0.27$, $p = 8 \times 10^{-4}$; theta $MD_{PT-RT} = 0.35$, $p = 8 \times 10^{-4}$; alpha $MD_{PT-RT} = 0.27$, $p = 7 \times 10^{-4}$) and RT compared to OBS (delta $MD_{RT-OBS} = 0.06$, p = 0.018; theta $MD_{RT-OBS} = -0.04$, $p = 0.018$; alpha $MD_{RT-OBS} = -0.05$, $p = 0.016$). (**b**) Boxplots of AUC calculated from retrained BCGNet and OBS method averaged in the task relevant region from 0.25 s to 0.75 s post stimulus. Hinges represent first and third quartile and whiskers span the range of the data ($MD_{RT-OBS} = 0.018$, $p = 1 \times 10^{-4}$).

One general difficulty in assessing BCG suppression quality is that the ground truth BCG signal is unknown. Unlike other current methods, the core of our method is designed to directly generate BCG from ECG. It is therefore possible to see how BCGNet might be used to simulate ground truth BCG signal, which can be augmented with simulated 1/f noise and brain derived sources in order to study the relative efficacy of other BCG suppression techniques. Furthermore, via extensions of BCGNet that either model the propagation of the ECG signal itself, or by directly injecting signal into the small central dense layer of the network, it may be possible to gain fine grained control of the BCG construction under test, for example, to study the different methods under changing heart-rate conditions.

One of the main advantages of our method is that because BCGNet is a software based solution, it can be used directly with existing commercial hardware, and does not require specially designed equipment. We hypothesize that our proposed method has scope for improvement as a wider body of EEG-fMRI data becomes openly available. As previously mentioned, we also consider that with the appropriate sensing equipment BCGNet is extensible to removing a broader array of artifacts, for example Electrooculography (EOG) or respiration artifacts. With the appropriate reference signal, it may also be useful in compensating for the helium pump artifact so that the Helium pump can remain on during EEG-fMRI scans [62, 68].

### C. Real-Time Applications

Hardware reference based methods [62], as well as some newer software based methods [28] do not rely on QRS complex detection. Methods that rely on QRS complex detection are likely to be difficult to extend to real-time BCG suppression because, as previously mentioned, QRS detection at high field strengths is difficult, and removal of the BCG around an upcoming QRS complex is likely to require a prediction of that QRS complex time. Small errors in that prediction are likely to result in poor BCG suppression. On the other hand, similarly to hardware based methods an RNN based approach allows us to implicitly estimate the instantaneous state of cardiac activity, and make use of it directly. This avoids the explicit prediction of the upcoming QRS complex, and likely makes us robust to deviations in heart rate which may occur over the course of an experimental session. A disadvantage that our method has in real-time applications is that it needs to learn the ECG-BCG relationship for each subject, so a training step must be performed before the main experiment.

## V. CONCLUSION

We have developed a deep learning based method for direct modeling of the BCG from the ECG, which can be used to reduce BCG related artifact in EEG-fMRI recordings, beyond the commonly used standard without the use of additional hardware. This method has the potential to be extended for use in validation of other methods, and in real-time applications. We believe these are encouraging results that are likely to be improved upon with more data, and the merging with developing hardware methods to make use of simple reference signals to address artifacts that we could not cater for in the current iteration of BCGNet.

ACKNOWLEDGMENT

We thank Anna Jasper for comments. We acknowledge computing resources from Columbia University's Shared Research Computing Facility project, which is supported by NIH Research Facility Improvement Grant 1G20RR030893-01, and associated funds from the New York State Empire State Development, Division of Science Technology and Innovation (NYSTAR) Contract C090171, both awarded April 15, 2010.

# Supplementary Materials

*Classification on band amplitude*

Data was processed as described in our task relevant single-trial classification analysis section, however prior to epoching, the BCG suppressed signal from each method was converted to a band amplitude measure. This conversion entailed initially filtering the data with an FIR band pass filter, with cutoffs matching either the delta, theta or alpha band. We then applied the Hilbert transform, took the absolute value of the result and Z-scored across the entire time series.

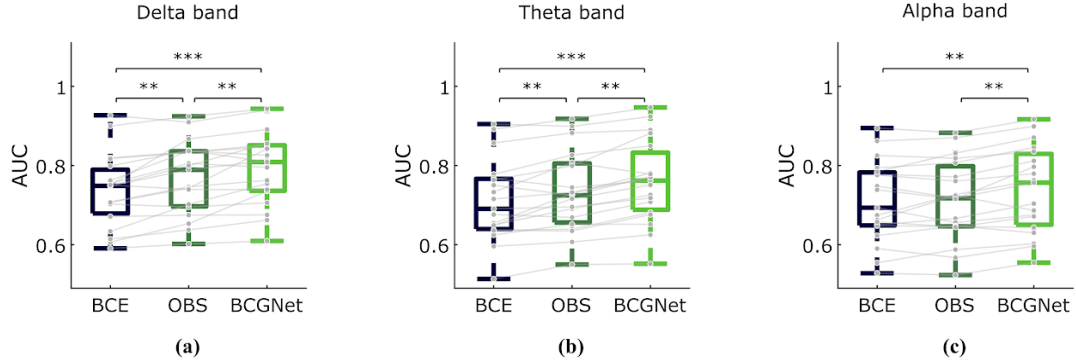

Supp. Fig. 1. Overall classification quality based on band amplitude is consistently improved by application of BCGNet when compared to OBS. (**a**) Boxplots of AUC averaged in the task relevant region from 0.25s to 0.75s post stimulus for delta band amplitude. Hinges represent first and third quartile and whiskers span the range of the data ($MD_{BCE-OBS}$ = -0.036, $p$ = 0.002; $MD_{BCE-BCGNet}$ = -0.044, $p$ = 8 × 10$^{-4}$; $MD_{OBS-BCGNet}$ = -0.019, $p$ = 0.005). (**b**) As (a) for Theta band amplitude ($MD_{BCE-OBS}$ = -0.025, $p$ = 0.007; $MD_{BCE-BCGNet}$ = -0.040, $p$ = 5 × 10$^{-4}$; $MD_{OBS-BCGNet}$ = -0.029, $p$ = 0.001). (**c**) As (a) for Alpha band amplitude ($MD_{BCE-OBS}$ = 0.005, $p$ = 0.717; $MD_{BCE-BCGNet}$ = -0.023, $p$ = 0.002; $MD_{OBS-BCGNet}$ = -0.029, $p$ = 0.004).

*Across-subject BCGNet compared to within-subject BCGNet*

While as shown in Fig. 8a BCGNet outperforms OBS after within-subject retraining, performance is worse, when compared to direct within-subject retraining (Supp. Fig. 2a). However the number of training epochs required to reach a local-minima during training is substantially decreased (Supp. Fig. 2b).

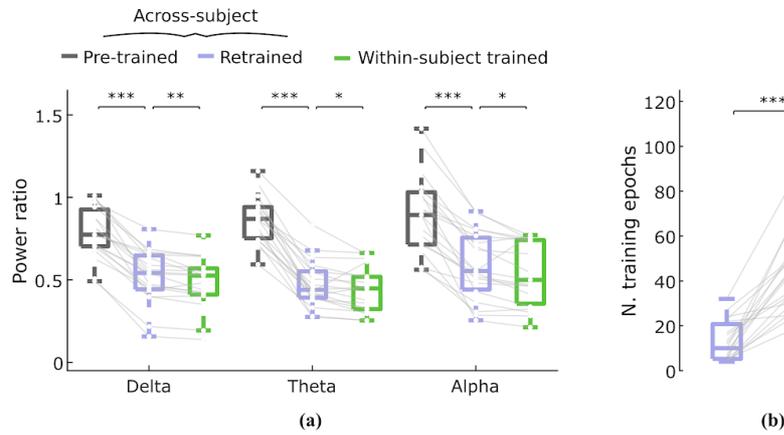

Supp. Fig. 2. Transfer in network trained across subjects. (**a**) Power ratio for BCGNet trained on all subjects and evaluated on held-out subjects for average channel (gray, pre-trained, PT) demonstrates poor performance, which recovers to levels near those of within-subject trained networks (green, as in Fig. 4, within-subject, WS) after training on the held out subjects (blue, re-trained, RT). Bonferroni-holm correction was carried out across bands and networks, although we limited our tests to PT compared to RE (delta $MD_{PT-RE}$ = 0.27, $p$ = 8 × 10$^{-4}$; theta $MD_{PT-RE}$ = 0.35, $p$ = 8 × 10$^{-4}$; alpha $MD_{PT-RE}$ = 0.27, $p$ = 7 × 10$^{-4}$) and RE compared to WS (delta $MD_{RE-WS}$ = 0.02, $p$ = 0.004; theta $MD_{RE-WS}$ = 0.03, $p$ = 0.014; alpha $MD_{RE-WS}$ = 0.05, $p$ = 0.014). (**b**) A pre-trained network reaches minimum validation MSE considerably faster than direct within subject training (re-trained network median: 10, within-subject trained network median: 59, $p$ = 1 × 10$^{-4}$).